# An IoT-based Intelligent Wound Monitoring System

**Hina Sattar, Imran Sarwar Bajwa, Nadeem Sarwar, Umar Shafi, Noreen Jamil, M.G Abbas Malik**

Corresponding author: Imran Sarwar Bajwa (e-mail: isbajwa@cs.bham.ac.uk).

This paragraph of the first footnote will contain support information, including sponsor and financial support acknowledgment. For example, "This work was supported in part by the U.S. Department of Commerce under Grant BS123456."

**ABSTRACT**: Clinical research of wound assessment focused on physical appearance of wound i.e. wound width, shape, color etc. Although, wound appearance is most crucial factors to influence healing process. however, apart from wound appearance other factors also contribute in healing process. Wound internal and external environment is one such factor that may show positive or negative impact on healing. Internet of things extensively popular during last decade, due to its heavy applications in almost all domains i.e. agriculture, health, marketing, banking, home etc. Therefore, in current research we proposed IoT based intelligent wound assessment system, for assessment of wound status and apply entropy and information gain statistics of decision tree to reflect status of wound assessment by categorization of assessment results in one of three class i.e. good, satisfactory or alarming. We implemented decision tree in MATLAB, in which we select ID3 algorithm for decision tree which based on entropy and information gain for the selection of best feature to split the tree. The efficient feature split of decision tree improved training accuracy rate and performance of decision tree.

## I. INTRODUCTION

Wound assessment is important healthcare concern in medical and clinical research. There is strong justification for regular wound inspection i.e. wound assessment is necessary to monitor treatment outcome, infection identification and evaluation of treatment accuracy. Clinical researcher believed that wound monitoring proficiently based on histological tracking of the morphological changes in tissues. Therefore, comprehensive wound assessment need regular clinical observation [1,2]. But with technology inventions of current era it became emerging need to convert clinical centric wound health care to patient centric health care by providing IoT based health care application composed of sensors, device, analysis modules and patients. [3]. Therefore, many current researches by IT Researchers focused on factors effecting wound healing and emerged with such solution for assessment of wound by measuring physical characteristics of wound e.g. wound size, wound color using IoT. [4-9].

Although, measurement of wound visual appearance play vital role in wound assessment, but there are many other internal and external wound factors which may have major impact on wound healing i.e. hydration level, skin temperature, oxygen saturation [1], environmental factors (air temperature, air humidity, presence of microbes, dust etc.) [10]. Therefore, wound care domain required such wound monitoring system that may also consider wound internal and external factors for wounds assessment, rather than just wound appearance. In health care domain there are many applications which designed to monitor different diseases but wound care domain still required such techniques and sensing system which can identify different factors of wound area including temperature, blood pressure, oxygen and infection status of wound using biosensor [11].

In healthcare domain researchers proposed a lot of wearable devices for health centric task i.e. disease diagnosis by taking patient body parameters as input, these wearable devices having biggest limitation that they were only used for disease diagnosis and unable to read data from patient body therefore these devices unable monitor disease condition, drug response by analysis of real time patient body characteristic [12].A lot of IoT based health care solutions based on sensors used for detection of medical conditions, efficiently with bearing low cost and easy access, however they may face issues due to variations in sensor data and poor data analysis techniques. [13]. Furthermore, there are many health care applications for wounds by using IoT and different decision making approaches [14-18], as shown in Table 1, structural and working description of these approaches [4,12,13,14-18] provided us given research goals.

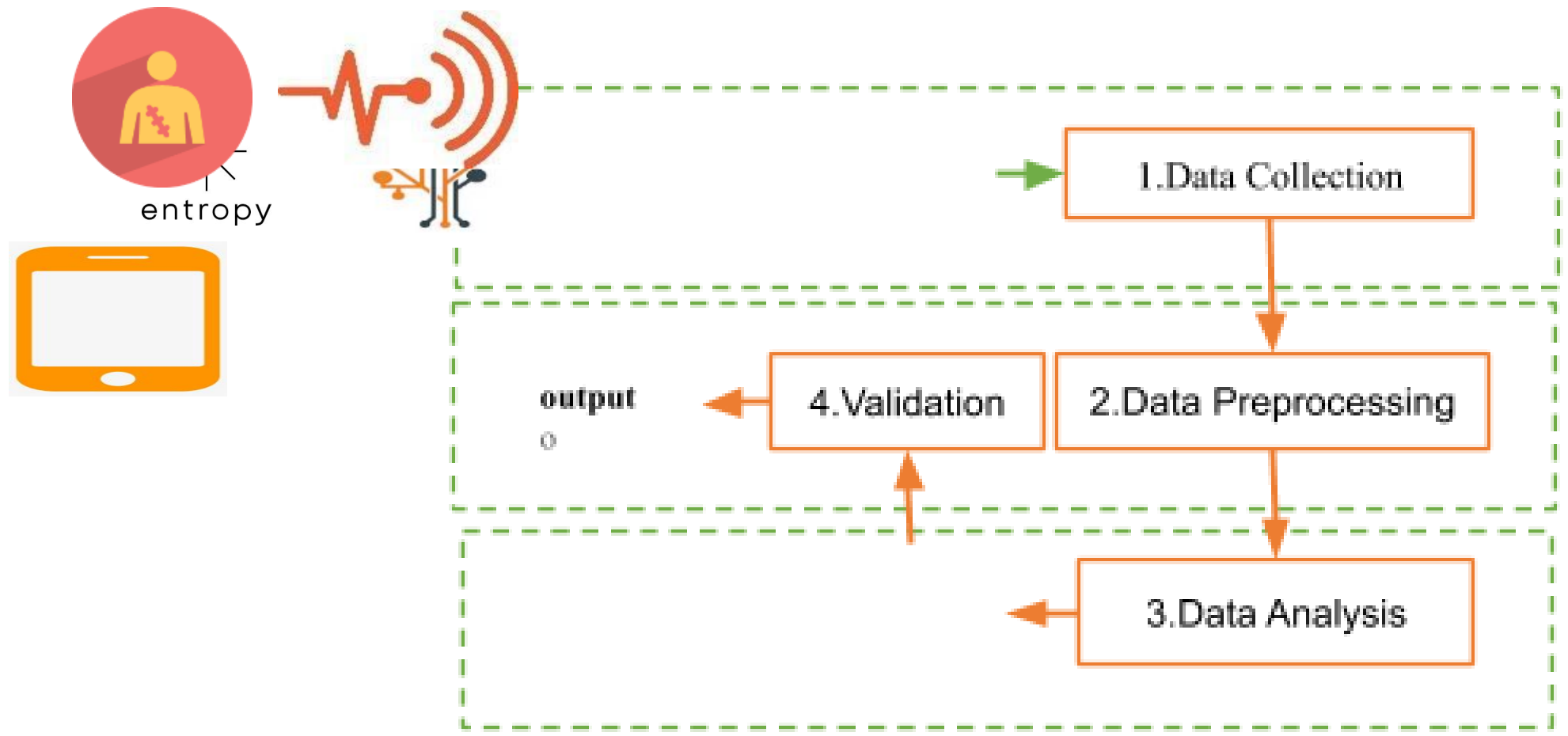

FIGURE 1: Wound Assessment System Model.

--Previous health care solution take input from patients and unable to read real time patient body conditions for analysis and prediction of health concerns.

--Previous health care solution used expensive and complex sensors in wearable devices for detection of disease diagnosis.

--Previous wound care solutions replicate clinical research to measured wound physical appearance for wound assessment by taking images of wound using smart phone camera.

--Previous wound care solution didn't use entropy methods for analysis of wound factors in wound assessment task.

--Previous wound care solution usually measured just single wound factor by sensor and analyze results based on single factor value.

In current research work we focused on wound assessment by measuring and analyzing internal and external characteristics of wound e.g. skin temperature, air humidity, air temperature, body oxygen saturation. We designed an IoT based intelligent wound assessment system to read real time wound internal and external factors and apply efficient entropy based decision trees for analysis and interpretation of current wound condition based on measured factor values. Our proposed system exhibit given features.

--Unlike traditional clinical research, our proposed system focus on real time measurement of important wound internal and external factors by sensors for effective wound assessment.

--Our proposed system used low cost hardware components, which ultimately provide patient with affordable wound assessment system.

--Our proposed system used entropy methods in decision tress, making decision trees more efficient by using key feature based split criteria.

--Our proposed system facilitate patient to do wound assessment at their home and consult with health experts immediately in case of emergency.

--Proposed System provide cost efficient wound assessment where no physicians or medical expert is involved, patient can track wound healing and would need to consult with medical expert only in case of emergency.

--Intelligent wound assessment at home save patient time as patient did not need to regularly visit medical expert for wound assessment.

Basic design of our proposed system as depicted in Figure 1, consists four major parts i.e. data collection by sensing system, data normalization, data analysis by entropy based decision trees and output interpretation.

The rest of the paper is structured, as follows: Section 2 discusses state of the art techniques and methods for management of wound healing; Section 3 describes the architecture of proposed intelligent wound assessment system, used hardware details and data analysis technique.; Section 4 provides details of the experiments their results and discussions to show the performance testing ,outcomes and limitations of the presented approach; and, Section 5 presents a conclusion and future work of the presented research.

TABLE 1:
STATE OF ART FEATURES AND LIMITATIONS

| Work Year | Sensor /Data Mining | Purpose | Limitation |
|---|---|---|---|
| [14] 2008 | RFID, accelerometers, decision trees | Elder care | Fix movement detection<br>Complex configuration<br>Expensive setup |
| [15] 2010 | 4 accelerometers, decision trees | Human activity detection | |
| [16] 2012 | Array of gas sensor | Odor detection for bacteria identification in wound | Helps to identify wound bacteria not able to protect from bacteria |
| [17] 2015 | Wireless and oxygen sensor | Wound monitoring bandage | Consider only one factor for monitoring |
| [18] 2017 | Temp, Pressure, Heart Rate sensors/ C4.5 Decision Trees | Health care monitoring | Facilitate specific patients not for all |
| [19] 2018 | Cloud of IoT devices/decision trees, KNN | eHealth Monitoring Framework for students | Expensive Technology<br>Not an application |
| [20] 2018 | Cloud of IoT devices/decision trees, KNN | eHealth Monitoring Framework | Expensive Technology<br>Not an application |
| [21] 2018 | Uric acid Biosensor | monitor wound healing | expensive biosensors |

## II. STATE OF THE ART

In health care domain, many researchers proposed easy solutions to facilitate patient in wound assessment efficiently by adopting different methods. In current section, we briefly describe research literature of health care domain using IoT and data mining approaches. We shortly described features of previously adopted method, their goals and limitation in Table 1, to highlight need of currently proposed wound assessment system.

Yu-Jin Hong et al. [14] proposed IoT based wearable sensors system for elder care. Their proposed system used accelerometers and rfid sensors to detect daily activities of elder. They used decision tree based classification to detect 5 human body states, along with detection of RFID tagged objects with hand movements.

Hyung-Gi Byun, et al. [15] proposed odor detection system for detection of bacteria type at early stage of wound infection. Their proposed system used an array of gas sensor. They used clinical samples collected from patients and used laboratory analysis techniques to differentiate infected and uninfected patients.

Wallace Ugulino et al. [16] proposed human activity recognition system HAR. Their proposed system used wearable device composed of 4 accelerometers for data collection and they provided dataset comprises 165,633 samples and 5 classes by applying classification with decision trees.

Pooria Mostafalu et al. [17] proposed smart bandage for wound monitoring based on wireless and oxygen sensors. They created 3D printed smart dressing systems for wound. Their proposed system indicates oxygen concentration which is important factor to track wound healing. Their proposed bandage facilitates monitoring of wound by wearing smart bandage at home.

V.Santhi et al. [18] proposed IoT based wearable device for monitoring health of pregnant ladies by reading environmental factors e.g. pressure, temperature and patient heartbeat. In case of abnormal factors their device will generate alarm and communicate with webApp using Wi-Fi. They used C4.5 decision trees classification algorithm for output prediction.

Prabal Verma et al. [19,20] proposed a cloud-centric IoT based smart m-healthcare monitoring framework for students. Their proposed framework predicts disease severity level by comparison with health measurement taken from IoT devices and medical domain. They used dataset of 182 suspected students and applied different classification techniques for pattern matching with K-cross validation. They compare performance of classification algorithms by comparing their accuracy, sensitivity and response time and concluded that decision trees and K-nearest neighbor outperforms other algorithms.
Sohini Roy Choudhury et al. [21] design wearable device to monitor wound healing with help of uric acid biosensor. Their proposed device detects Uric Acid(UA) from wound. UA termed as biomarker, which have strong correlation with wounds and their healing. They used to be doxelectron shuttle, ferrocene carboxylic acid(FCA) which made transfer of electron between the enzyme and the transducer. For Uric acid detection in wound fluid they used wound fluid volume range as 0.5–50μL. Their case Studies from different wound samples have shown an average recovery of 107%.

## III. Architecture of Proposed IWAS:

Our proposed Intelligent wound assessment system IWAS used IoT approach for measurement of real time wound factors i.e. Air temperature and humidity, body oxygen saturation and body temperature and predict effect of these factor on wound healing by doing classification with decision trees. Proposed system applied Entropy, information gain statistics to choose best split criteria for decision tree. Architecture of proposed approach shown in Figure 2, in which we depicted all components, their interaction and working strategy of proposed IWAS graphically.

FIGURE 2: Architecture of IWAS.

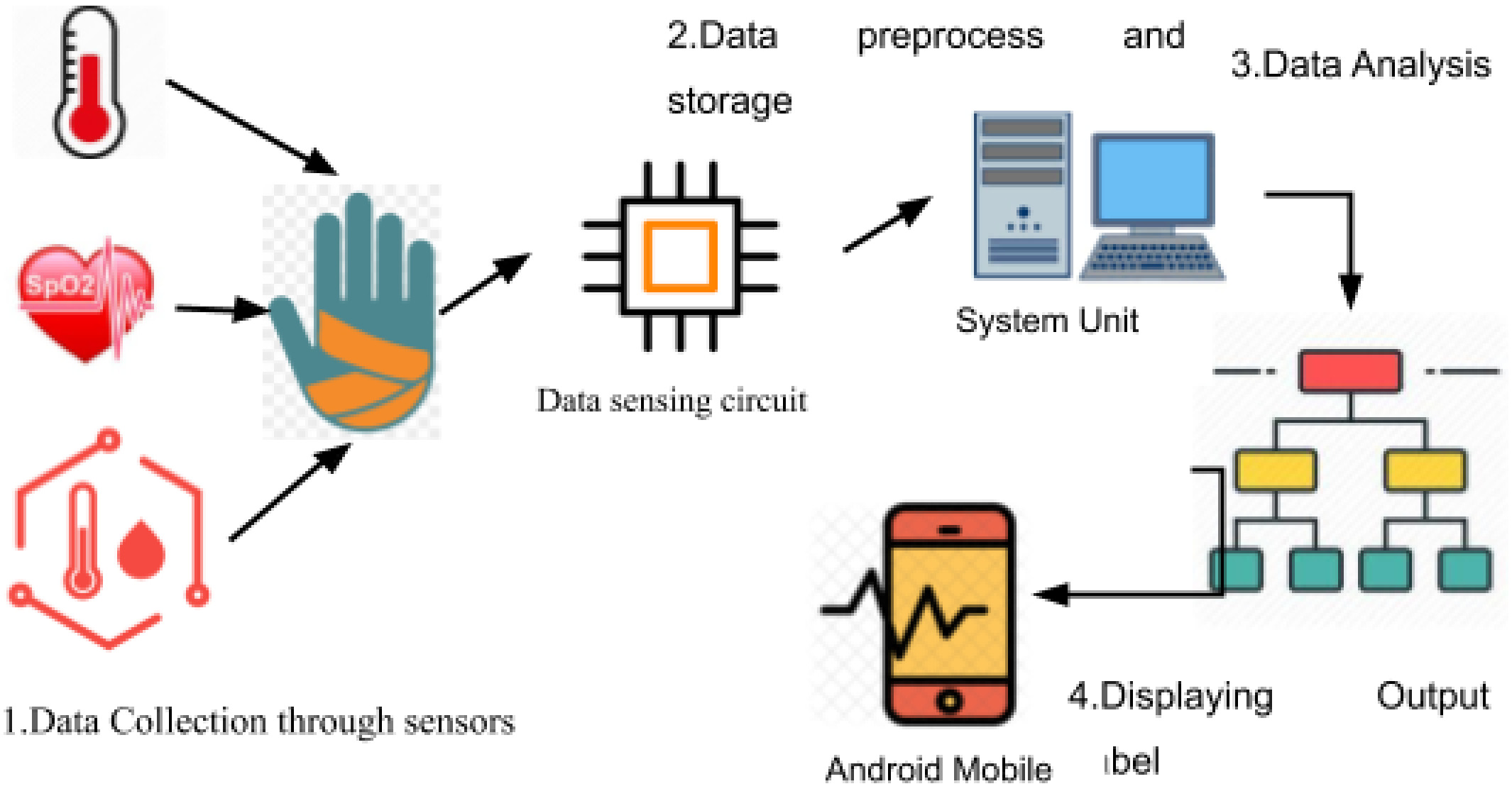

### *A. Used Algorithm*

We designed comprehensive working algorithm for Intelligent wound assessment system based on entropy methods of decision tree. Our proposed algorithm steps described in given Algorithm 1.

| Algorithm 1: Working algorithm of intelligent wound assessment system (IWAS) | |
|---|---|
| Step 1 | Data collection is first step of proposed system for which IWAS used its first component RSS. |
| Step 2 | After data collection next step is data preprocessing to normalize data of each interval and obtained values will be stored in dataset. |
| Step 3 | After data preprocessing ,next step is data analysis which for which we used Decision tree classifier by applying entropy ,information gain statistics to choose best split criteria. |
| Step 4 | In last step ,computed output match with expected one for validation of system performance. |
| Step 5 | In Last step system will display predicted result to targeted user. |

### B. Used Hardware

IoT based system used embedded sensing devices to efficiently and economically sense and record real time data [22,23,24]. We also used efficient sensors for real time data sensing. Our proposed IWAS, used sensors based circuit for real time reading of wound internal and external factors, which then provided to decision trees for analysis of healing after preprocessing. The hardware components involve in data collection circuit depicted below.

Arduino UNO Microcontroller

--DHT22 temperature and humidity Sensor

--LM35 Temperature sensor

--MAX30100 Heart Rate Module

Our designed circuit used above hardware components. We used three sensors DHT22 for measurement of air temperature and humidity, LM35 for measurement of skin temperature and MAX301000 heart rate module to measure oxygen saturation in blood also termed as SpO2. We connected these sensors with used microcontroller i.e. Arduino UNO to make circuit for effective data collection.

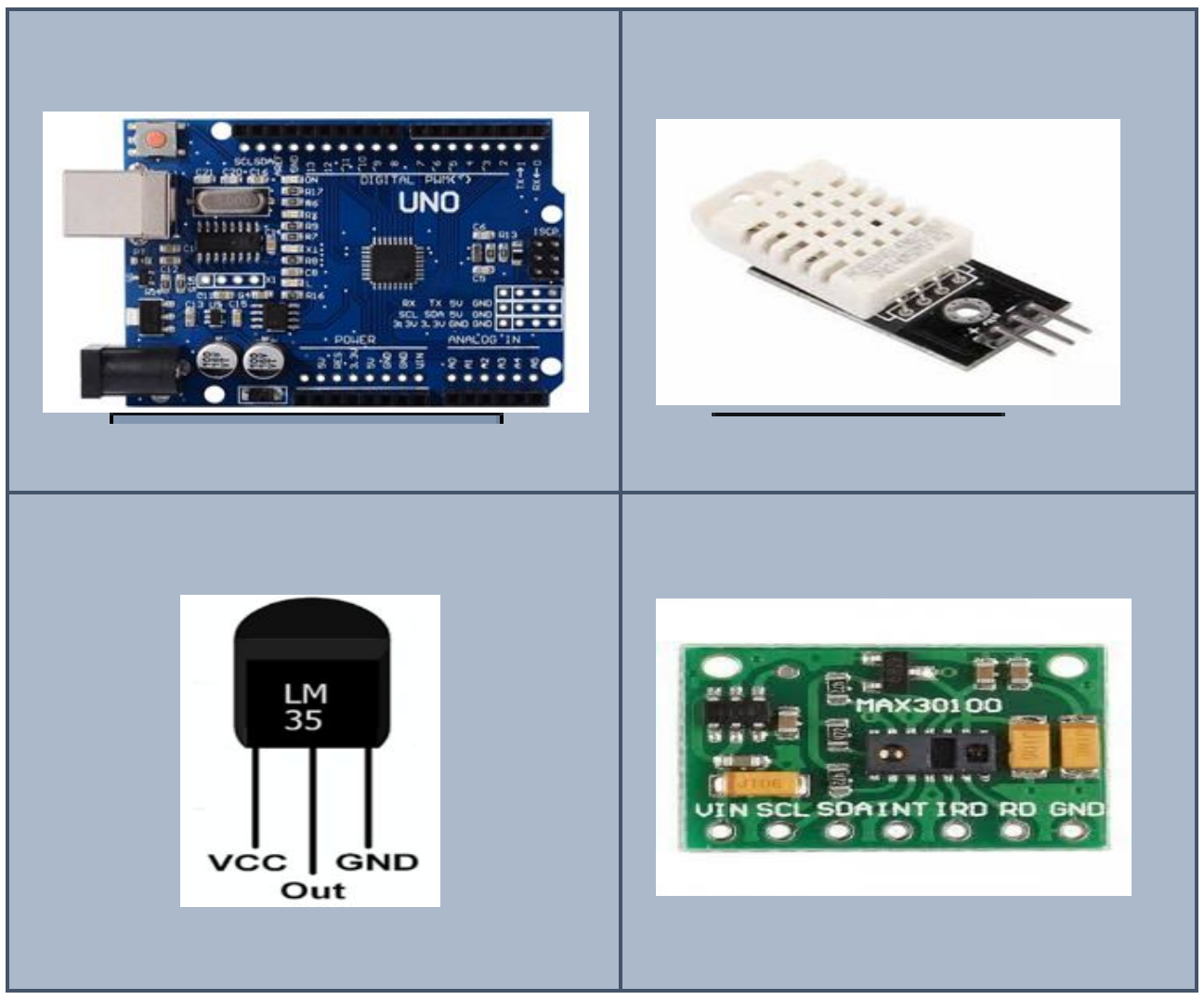

**FIGURE 3. Hardware components of IWAS.**

### C. Used Analysis Technique

In our proposed IWAS, factors read by Arduino based circuit used for analysis, to predict their impact on wound healing. There are many available techniques in machine learning for analysis of data to predict outcome e.g. SVM, Neural Network, KNN, Random Forest, Decision Tree. There are many applications of classification algorithms for providing solutions of health care concerns e.g. Vikas Chaurasia et al. [25] used classification techniques to predict breast cancer type. Different researchers applied different techniques based upon problem scenario and expected result. In our proposed system we used decision tree for prediction of wound factors (under consideration) effect on wound healing. We preferred decision tree for data analysis due to following reasons also given by [26,27,30].

We select Decision trees for analysis as it is suitable algorithm to work with problem scenario in which input is in form of attribute-value pair as our scenario temperature-hold, mild, cold.

We preferred decision trees over other classification algorithm as it can deal with error of training data by pruning techniques.

Decision tree uses different measures such as Entropy, Gini index, Information gain etc.to find best split of attributes.

#### 1) Decision Tree Structure

In our proposed approach we performed data analysis step by using decision tree. The decision tree could use to solve problem of classification or regression. In our problem scenario we used decision tree to perform classification on test data set. Decision tree used for supervised learning problems to predict value of dependent variable from local region of input space. Decision tree can simply define as a graph G= (V, E) having finite non empty set of nodes, where V= nodes and E=edges between nodes. [28,29]. These nodes are of three types i.e. root, leaf, internal. Decision tree starts from root node, each internal node put condition on attribute value of training data set and leaf node give predicted class value of attribute value based on condition result. Decision tree construction follow divide and conquer rule where each path builds a decision rule. [30,31].

In our proposed approach we selected ID3 algorithm for decision tree, consider decision tree containing three categories of wound assessment. We divided wound healing impacts in three classes good, satisfactory, alarming. $\alpha^{+}$ represent all instances of good impact class, $\alpha$ represent all instances of satisfactory impact class, and $\alpha^{-}$ represent instances of all alarming impact class where $\alpha^{=}$ $\alpha^{+} + \alpha + \alpha^{-}$ represents all instances of dataset. For an instance of input data set Xi (i Ɛ α), Yi denotes classification scores/output of target category. [29]. We induced decision tree from provided training data set in top down manner for each input instance. Where each input instance contains K attributes, in current scenario K1=wound temperature, K2= oxygen saturation, k3=air temperature and k4=air humidity. Every node in generated decision tree is associated with decision/splitting attribute which could selected by attribute selection strategy based on different algorithms and selection measures. Most commonly used learning algorithms for decision trees are ID3, C4.5 and CART. [28]. We used ID3 algorithms for learning of decision tree which used entropy and information gain as an information measure for computation and quality evaluation of a node split by a given attribute. [32,33].

### 2) Used ID3 Decision Tree Algorithm

In our proposed system, we implemented decision tree in MATLAB, by using top-down induction approach for tree generation. MATLAB used Hunt's induction algorithm for decision tree generation, consisting given steps [28].

In our proposed DT, Let Xt be the set of training instances for node t and y = {y1, y2..., yk} be the class labels

t is leaf node if all the instances in Xt belong to the same class yt

If instances in Xt belongs to more than one class, a test condition will be designed for partitioning off instances into smaller subsets. Then for each outcome of test condition child node is created and the instances in Xt are distributed to the children based on the outcomes.

Recursively apply the algorithm to each child node.

We used ID3 algorithm given below to implement decision tree, working steps of ID3 also depicted in flowchart give in Figure 4, also used by [34].

| Algorithm 2: ID3 Working Algorithm of Decision Tree | |
|---|---|
| Step 1 | Calculate entropy of every feature F in input dataset S. |
| Step 2 | Calculate information gain of every attribute F using its entropy to find one with highest gain and minimum Entropy. |
| Step 3 | Divide dataset S into subsets using the Feature F for which the entropy after splitting is minimized, in other words whose information gain is maximum. |
| Step 4 | Make node of decision tree which contains that Feature. |
| Step 5 | Repeat step 2 to 4 for remaining Features. |

Test condition is applied on attribute for splitting node in subset, key factor in decision tree performance is selection of such attribute which could better discriminate input data. In our used Decision tree, we adopted Information theory based criteria known as **information gain,** for measurement of attribute in order to choose best one for node split. Data pick up is itself judged by measure called **entropy**, which utilized to measures the impurity of set of training objects. For a collection of data set S, entropy formula is given in equation 1, formulated by Claude Shannon, also used by [28,35-38]. Entropy helps decision tree to determine how informative a node is. [32]

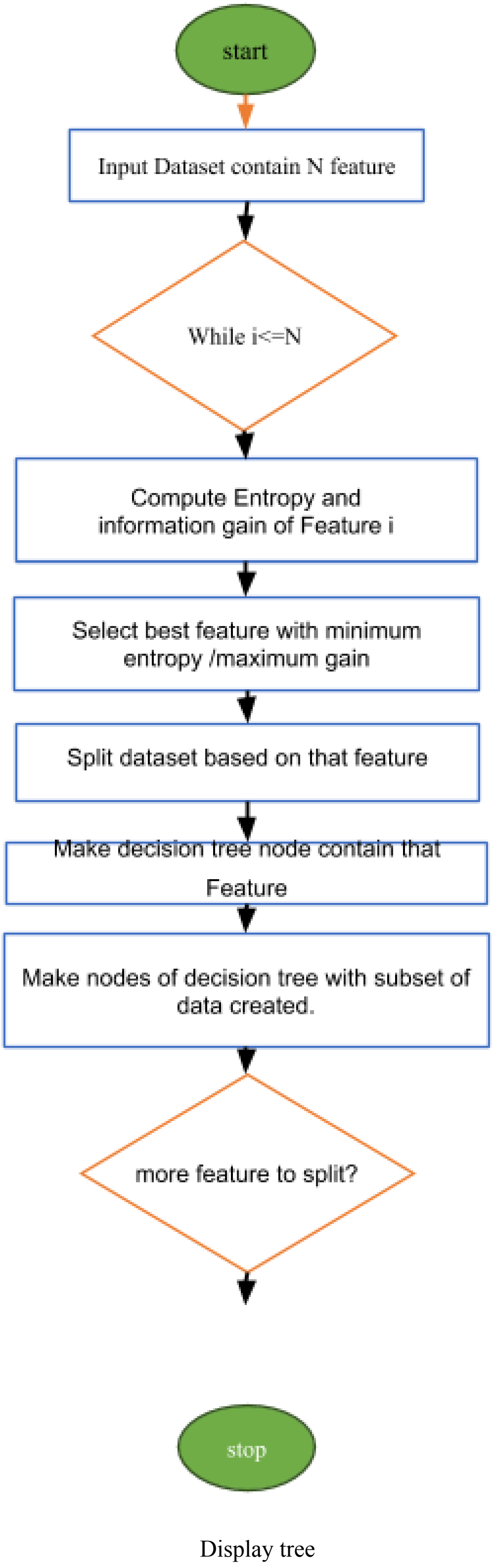

**FIGURE 4. Flowchart of ID3 Decision Tree**

$$Entropy\,(s) = \sum_{i=1}^{c} - pi \log log\ 2pi \qquad (1)$$

Where pi= simply probability of and element/class in our sample data.
Best split could have obtained by applying information gain, which is simply amount of information obtained by knowing the value of the attribute i.e. the entropy of the distribution before the split minus the entropy of the distribution after it. The largest information gain is equivalent to the smallest entropy. Simply, Information gain use entropy to identify what attribute is best to select for tree split [32]. We used formula for calculation of information gain given in equation 2 and 3 also used by [39].

***on gain***= (Entropy of distribution before the split)– (entropy of distribution after it). (2)

$$) = Entropy(T) - Entropy(T, X) \quad (3)$$

## IV. Implementation

In previous section, we discussed complete design of proposed IWAS. There are two major components of IWAS i.e. sensing part and analyzing part. We implemented sensing part by ARDUINO and analyzing part in MATLAB. In this section, we described implementation of IWAS comprehensively.

### A. *Measuring Wound Factors by Sensing System:*

First component of proposed IWAS is sensors based system which we used to measure different wound factors for wound assessment. We implemented this portion with Arduino based circuit which we designed by hardware components discussed in section 3.3. Sensing system collected wound factors and used standard values of body temperature, air temperature, air humidity and SpO2.

#### 1) Measuring Body Temperature

Temperature is major factor included in wound characteristics which boost/delay healing by directly and indirectly effecting other wound characteristics. Temperature can change chemical and enzymatic actions of healing process which in turn effect healing. It is accepted by Wound Care Education Institute that all enzymes and cells functions properly in normal body temperature. Any increase and decrease in temperature negatively affect the wound healing process. Wound healing effected by temperature as body temperature could affect local blood flow and lymphocyte extravasation, moreover temperature is early indicator of infection which determine wound chronicity. [40].
Therefore, we considered body temperature as important factor in our proposed wound assessment system. We used temperature standard ranges provided in Table 2, for measurement of body temperature.

Table 2.
Standard Body Temperature Range

| Standard | Celsius | Fahrenheit |
|---|---|---|
| Hypothermia | <35°C | 95.0°F |
| Normal | 36.5-37.5°C | 97.7-99.5°F |
| Fever/Hyperthermia | >37.5or 38.3°C | 99.5-100.9°F |
| Hyperpyrexia | >40.0 or 41.5°C | 104.0-106.7°F |

#### 2) Measuring Air Temperature

Air Temperature significantly affect human body temperature, if environmental temperature is higher than normal range it causes sweating, as a result skin loses its moisture and temperature of skin tissues drops. This may lead disturbance in human body temperature balance, which may negatively influence on wound healing. Therefore, we considered air temperature an important factor for measurement in our proposed IWAS. we used standard range given in Table 3, for measurement of air temperature.

Table 3.
Standard Temperature Range

| Class | Temperature Range |
|---|---|
| Normal (Winter) | 16 °C -18 °C |
| Normal(Summer) | 20 - 23.5 °C |

#### 3) Measuring Air Humidity

Air temperature feel have significant relation with air humidity level i.e. high level of air humidity rises temperature feel e.g. 90° temperature feel like 90 ° if humidity is 30% and may feel like 112° if humidity rises up to 65%. This high temperature feels cause sweating, which ultimately cool down body and skin lose its moisture therefore temperature of skin tissues drops.

Air Humidity also affect skin moisture level i.e. Low air humidity result in dry skin outermost layer called epidermis as skin water drawn out of the skin's surface into the air. While in presence of high air humidity skin moisture level sustains as body use its own natural moisturizing factors and it absorb water from the atmosphere to keep it hydrated.

Therefore, we considered air humidity important for wound assessment, we used standard humidity range given in Table 4, for measurement of air humidity.

TABLE 4.
STANDARD HUMIDITY RANGE

| Class | Humidity Range |
|---|---|
| Dry | 0-20% |
| Normal | 20%-60% |
| Wet | 60%-100% |

4) MEASURING OXYGENATION

Term oxygenation refers the oxygen saturation which means oxygen levels in blood. It is recognized widely that oxygen plays a significant role in all stages of wound healing. For all type of skin wound body need bacterial defense, cell proliferation, collagen synthesis and angiogenesis. British Journal of Dermatology reported that oxygen major role is its ability to produce energy. Howard M. Kimmel et al. [41] studied the effect of oxygen on wound healing and they define oxygen as key factor in wound healing. All cell on wound need energy for reproduction. Therefore, they need adequate amount of oxygen to generate energy. In absence of proper oxygen, a condition known as hypoxia occurs which can slow and even stop the healing process.

TABLE 5.
STANDARD OXYGEN RANGE

| **Standard** | **Oxygen %** |
|---|---|
| Normal | 95-100 |
| Hypoxemia | <95% |
| Higher | >100% |

TABLE 6.
WORKING RULES OF PROPOSED IWAS

| **Rule No** | **Air Temp** | **Air Humidity** | **Wound Temp** | **SpO2** | **Wound Assessment Class** |
|---|---|---|---|---|---|
| 1 | Normal | Normal | Normal | Normal | Good |
| 2 | High | Dry | Normal | Normal | Good |
| 3 | High | Normal | Normal | Normal | Good |
| 4 | Low | Normal | Normal | Normal | Good |
| 5 | Low | Wet | Normal | Normal | Good |
| 6 | High | Dry | Hyperthermia | Normal | Satisfactory |
| 7 | Low | Wet | Hyperthermia | Normal | Satisfactory |
| 8 | Low | Wet | Hyperthermia | Hypoxemia | Satisfactory |
| 9 | High | Dry | Normal | Hypoxemia | Satisfactory |
| 10 | Low | Wet | Normal | Higher | Satisfactory |
| 11 | High | Dry | Normal | Higher | Satisfactory |
| 12 | High | Wet | Hyperpyrexia | Normal | Alarming |
| 13 | Low | Dry | Hyperpyrexia | Normal | Alarming |
| 14 | High | Wet | Hyperpyrexia | Hypoxemia | Alarming |
| 15 | Low | Dry | Hyperpyrexia | Hypoxemia | Alarming |
| 16 | High | Wet | Hyperpyrexia | Higher | Alarming |
| 17 | Low | Dry | Hyperpyrexia | Higher | Alarming |
| 18 | Low | Dry | Hypothermia | Higher | Alarming |
| 19 | High | Wet | Hypothermia | Higher | Alarming |
| 20 | High | Wet | Hypothermia | Hypoxemia | Alarming |
| 21 | Low | Dry | Hypothermia | Hypoxemia | Alarming |
| 22 | Low | Dry | Hypothermia | Higher | Alarming |

Moreover, Oxygen is necessary for angiogenesis, which could define as normal process of new blood vessel formation from existing one. It is vital process for growth and development of body tissues and also essential part of wound healing process by growth of damaged tissues. Higher oxygen level play positive role in angiogenesis by increasing the rate and quality of new blood vessel growth
Correct collagen synthesis depends upon appropriate level of oxygen in blood.
Good Oxygen level increase blood vessel growth.
More oxygen in injured tissues facilitate more angiogenesis and with higher oxygen more collagen deposit. [42,43]
Therefore, in our proposed IWAS, we considered oxygenation level as important factor for assessment of wound. In order to analyze measured oxygen level we used given standards range provided in Table 5.

### B. *IWAS Working Rules*

Our proposed IWAS used decision trees for data analysis. Decision tree performed data analysis of input and concluded output based on patterns of data set which provided during training. This dataset patterns based on standard rules of system. Our proposed IWAS used working rules, given in Table 6, which we designed after observing standard ranges of wound factors (body temp, air temp, air humidity and oxygenation) given in Table2, 3,4 and 5 respectively and role of these factors on wound healing.

### C. *Implementing Decision Trees*

In our proposed IWAS, we implemented decision trees in MATLAB by using built in classification learner app feature of MATLAB. We depicted decision trees MATLAB implementation steps in flowchart, shown in Figure 5, consisting given simple steps.
Opening training dataset in MATLAB workspace.
Open classification learner app.
Import data set from MATLAB workspace.
Configure learner app.
Train decision trees on dataset provided.
Export trained decision tree on MATLAB workspace.
Use view command to display decision tree on MATLAB workspace.

#### 1) Used Training Dataset

We implemented proposed IWAS with Decision trees in MATLAB. In Figure 4 we depicted decision tree training process. We read training data set to MATLAB workspace as variable, which we further import as input data set during configuration of decision tree in classification learner app. Our used training data set comprises 650 input instances with pre-labeled wound assessment class. Structure of training data set shown in Table 7.

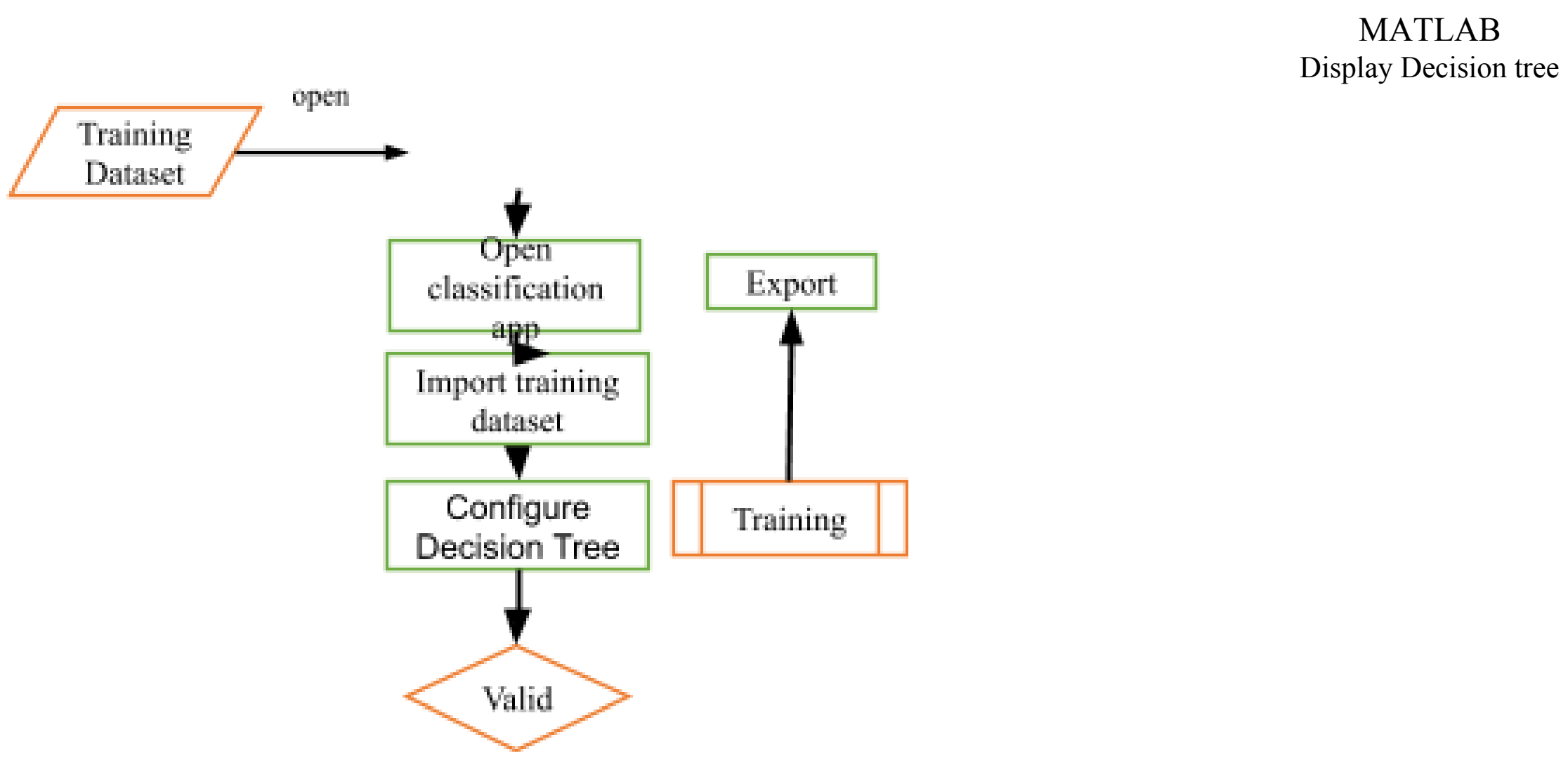

**FIGURE 5.** **Decision Tree Implementation Flowchart.**

TABLE 7.
TRAINING DATASET INSTANCES

| Class | Numeric Value | Total Instances |
|---|---|---|
| Good | 1 | 172 |
| Satisfactory | 0 | 84 |
| Alarming | -1 | 394 |

In our proposed training data set input instances had four features i.e. Wound temperature, air temperature, air humidity and SpO2, which we labeled with three output classes (good, satisfactory and alarming) by using rule set given in table 6. We provided comprehensive description of input instances count for each feature corresponding to each value of feature in Table 8.

### 2) TRAINED DECISION TREE

We implemented decision tree in MATLAB, by using built-in classification learner app. We first read training data set in MATLAB working space and import it in learner app after configuration of predictor and response variables i.e. in current system we had four predictor variables body temperature, air temperature, air humidity, SpO2 and one response variable assessment class which have three possible values 0,1 and -1 where

1=Good class

0=satisfactory class

-1=Alarming class

After configuration, we trained decision tree and export it on MATLAB workspace. The accuracy rate of trained decision tree is 96%. The trained decision tree of proposed IWAS shown in Figure 6. Decision Tree used entropy and information gain to split branches based on effective feature first. Our proposed decision tree used entropy and information gain formula provided in section III. In our proposed system, we calculated maximum value of entropy by using formula given in Equation 4.

$K$ (4)

where K= no of output categories, as there are three output categories in our proposed system so by putting values.

$K$ $3 = 1.5$

TABLE 8.
COMPREHENSIVE DESCRIPTION OF TRAINING DATASET

| Wound Temperature | | | | | Air Humidity | | | | |
|---|---|---|---|---|---|---|---|---|---|
| Assessment Class | Good | Satisfactory | Alarming | Total | Assessment Class | Good | Satisfactory | Alarming | Total |
| Normal | 172 | 63 | 33 | 268 | Dry | 20 | 40 | 79 | 139 |
| Hyperthermia | 0 | 21 | 81 | 102 | Normal | 126 | 02 | 136 | 264 |
| Hyperpyrexia | 0 | 0 | 117 | 117 | Wet | 26 | 42 | 179 | 247 |
| Hypoxemia | 0 | 0 | 163 | 163 | -- | -- | -- | -- | -- |
| Total | 172 | 84 | 394 | 650 | Total | 172 | 84 | 394 | 650 |
| Air Temperature | | | | | SpO2 | | | | |
| Low | 64 | 42 | 195 | 301 | Hypoxemia | 01 | -- | 149 | 150 |
| Normal | 87 | 42 | 119 | 248 | Normal | 170 | 24 | 61 | 255 |
| High | 21 | 0 | 80 | 101 | Higher | 01 | 60 | 184 | 245 |
| Total | 172 | 84 | 394 | 650 | Total | 172 | 84 | 394 | 650 |

So, the maximum value of entropy for proposed system should not exceed 1.5. Proposed system computed entropy of wound assessment and then compute entropy for each information split and computed information gain, the feature which showed maximum gain selected for tree split. We showed calculation of entropy and information gain for one feature i.e. wound temperature, proposed system used same computation strategy to calculate entropy and information gain of remaining three features and selected one with greater information gain i.e. wound temperature as shown in Figure 6.

Entropy (Wound Assessment)

$$=-\frac{172}{650}\frac{172}{650}-\frac{84}{650}\frac{84}{650}-\frac{394}{650}\frac{394}{650}\simeq 1.3$$

Entropy (Assessment/wound Temp=normal)

$$=-\frac{172}{268}\frac{172}{268}-\frac{63}{268}\frac{63}{268}-\frac{33}{268}\frac{33}{268}\simeq 1.20$$

Entropy(Assessment/woundTemp=Hyperthermia)

$$=-\frac{0}{102}\frac{0}{102}-\frac{21}{102}\frac{21}{102}-\frac{81}{102}\frac{81}{102}\simeq 0.73$$

Entropy(Assessment/woundTemp=Hyperpyrexia)

$$=-\frac{0}{117}\frac{0}{117}-\frac{0}{117}\frac{0}{117}-\frac{117}{117}\frac{117}{117}\simeq 0$$

Entropy(Assessment/woundTemp=Hypoxemia)

$$=-\frac{0}{163}\frac{0}{163}-\frac{0}{163}\frac{0}{163}-\frac{163}{163}\frac{163}{163}\simeq 0$$

Weighted Average

Entropy(Assessment/woundTemp)

$$=\frac{269}{650}*1.20+\frac{102}{650}*0.73+\frac{117}{650}*0+\frac{163}{650}*0\simeq 0.611$$

Information gain (assessment class, wound temperature)

Entropy(assessment)-Entropy (assessment/wound temp)

≈0.689

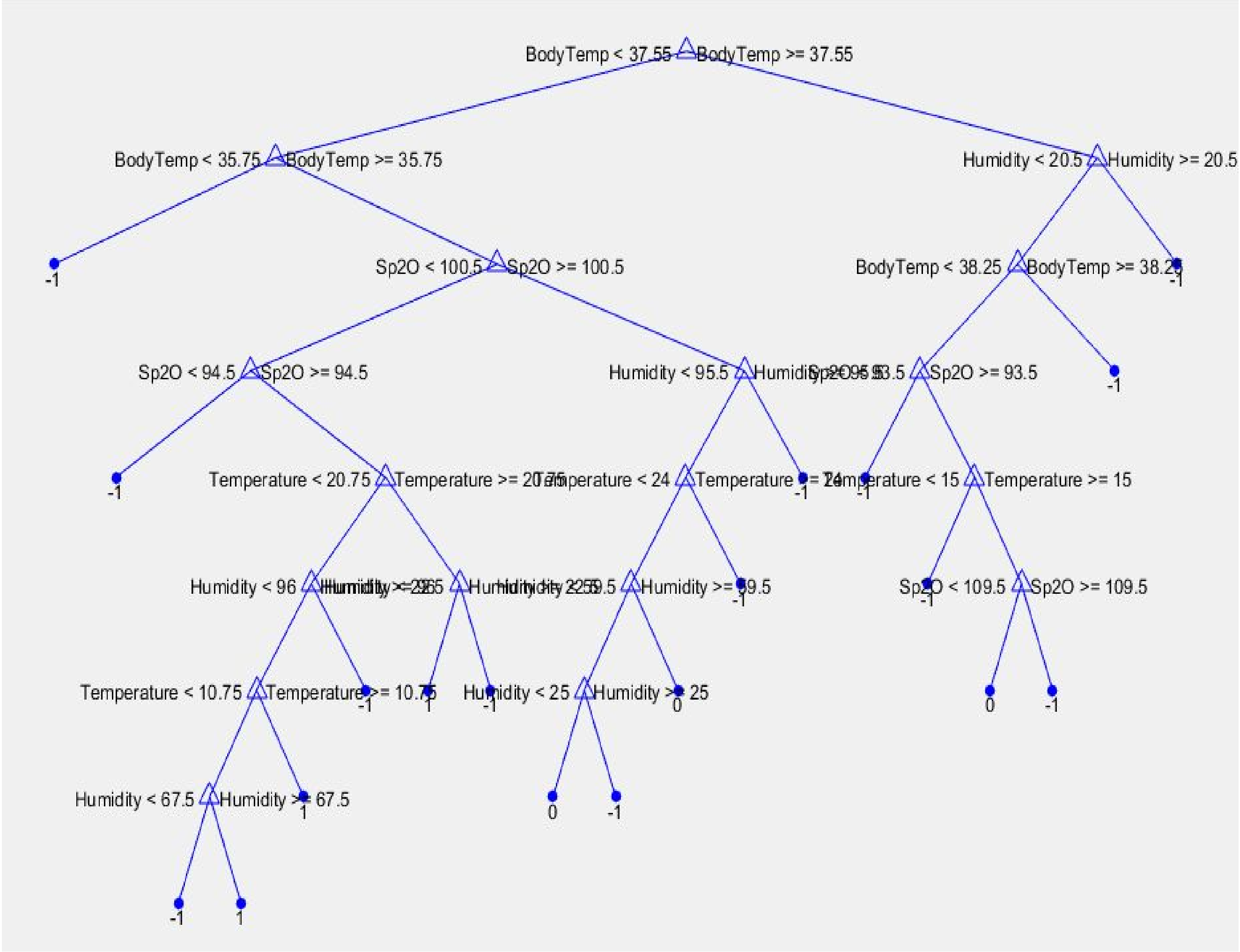

**FIGURE 6.** Trained Decision Tree View

## V. Experiments and Results

We used MATLAB to design and train decision tree, we depicted trained decision tree view in Figure 6s, In order to evaluate training process accuracy rate, we discussed three different measurement graphs in this section i.e. confusion matrix, ROC and scatter plot. We discussed performance rate achieved after training in this section and applied trained decision tree on different patient cases to test performance of decision tree after training.

### A. *Decision Tree Training Assessment*

Our proposed decision tree generated with 96.4% accuracy rate during training. Scatter plot between temperature and humidity shown in Figure 7, in which red dots showed truly classified instances of alarming class, blue dots showed truly classified instances of good class and green dots showed truly classified instances of satisfactory class. Similarly, misclassified instances showed by cross symbol red for alarming, blue for good and green for satisfactory respectively. Plot showed 4 cross for good class,6 for alarming class and 8 for satisfactory class, which showed that decision tree training process revealed less error rate.

We observed confusion matrix of our trained decision tree module as shown in Figure 8 to analyze its performance. Confusion matrix shown percentage along with count of truly predicted instances and falsely predicted instances of each output label, as it is shown in Figure 8 that out of 394 alarming instances 387 are correctly predicted and 7 instances are wrongly classified as satisfactory and good output label, while out of 84 satisfactory class prediction 71 are corrected and 13 are wrong, similarly out of 172 good label instances decision tree predicted 167 instances truly instances wrongly. Trained model of decision tree showed overall good percentage for all output labels i.e.98.2% for alarming,97.1% for good and 84.5% for satisfactory class labels

In Figure 9, we depicted ROC curve of our proposed trained decision tree. The ROC showed a true positive rate of the trained SVM on the y—axis and false positive on the X-axis. Area Under the Curve value of proposed decision tree = 0.9763.

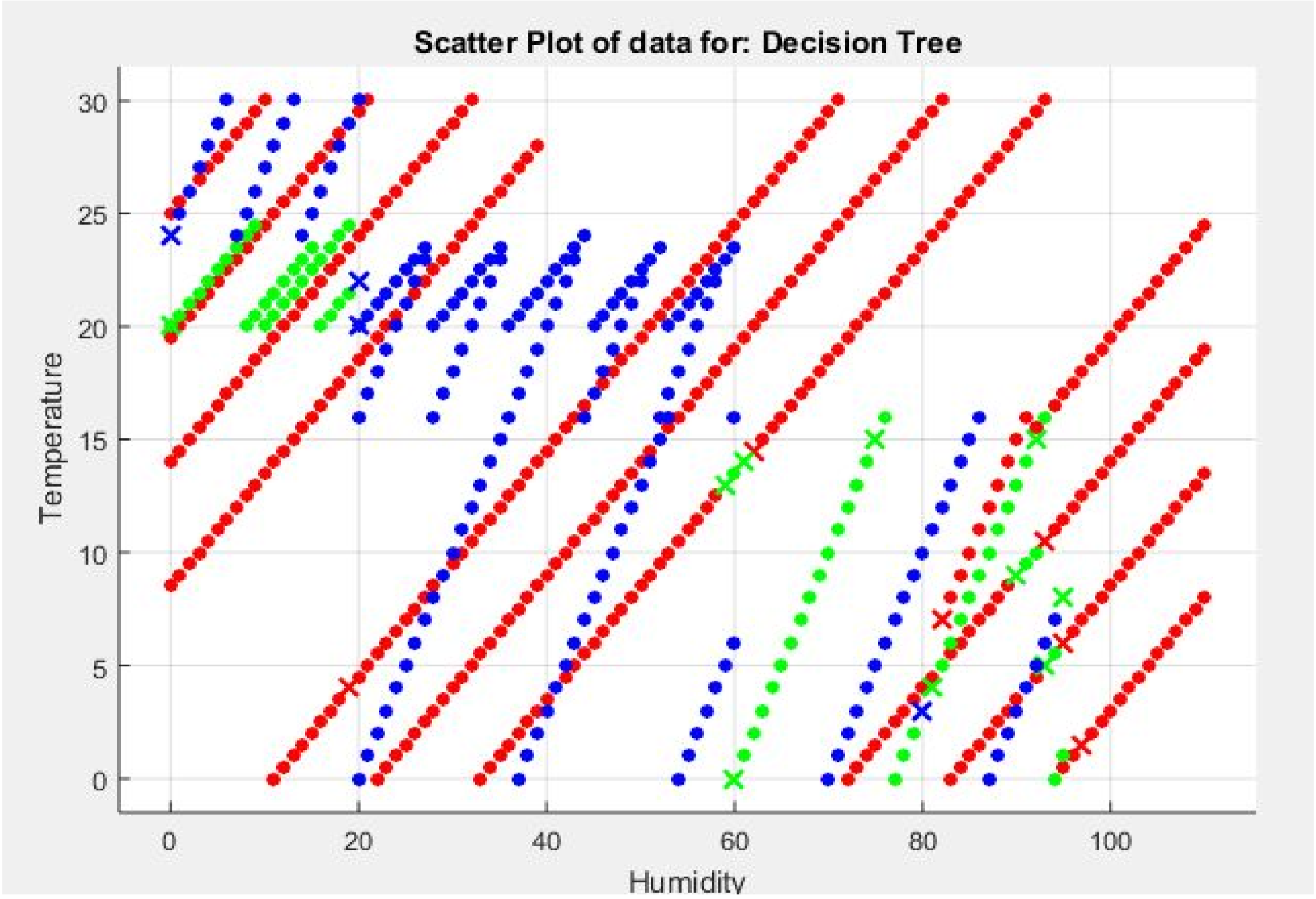

FIGURE 7. Scatter Plot of Decision Tree

B. *Decision Tree Classification*

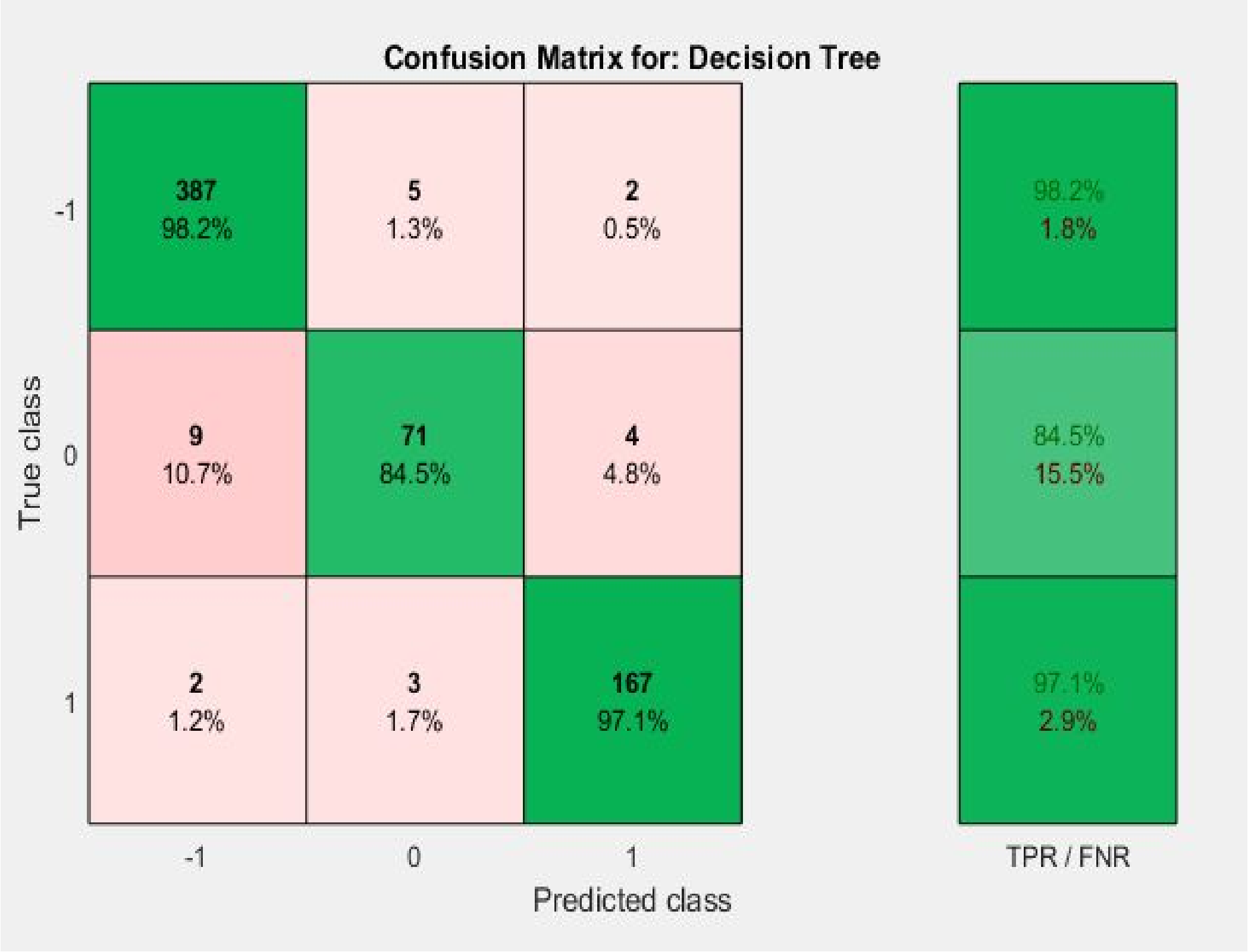

**FIGURE 8.** Confusion Matrix Of Decision Tree

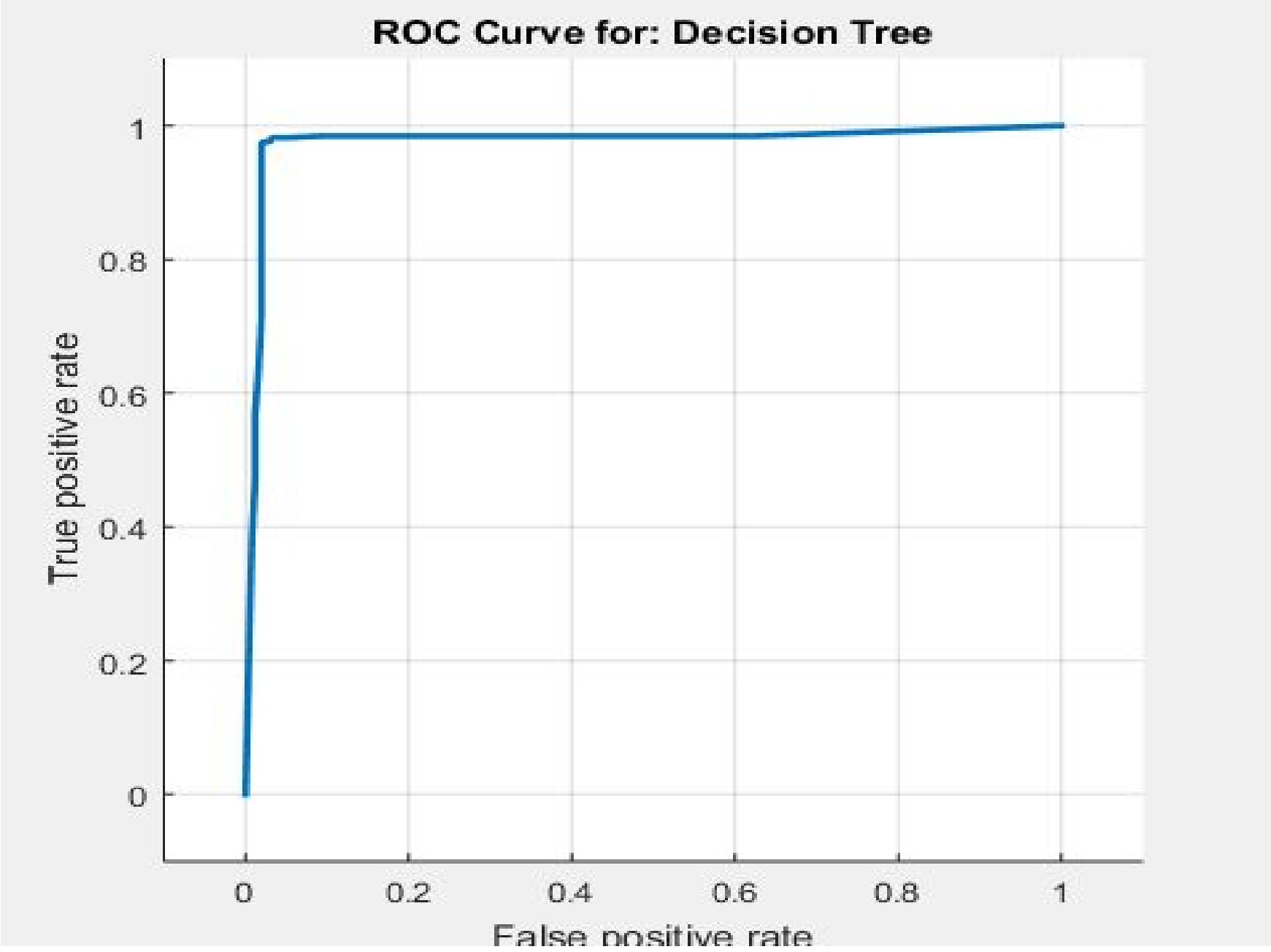

**FIGURE 9.** ROC Curve of Decision Tree

B. *Decision Tree Classification*

We did experiment to check performance of proposed IWAS. We collected 500 different input cases and provided as input to proposed system in order to check accuracy rate of proposed IWAS. We collected input data from 5 different patient suffering from skin wounds by using proposed sensing component of IWAS provided these input cases to analysis part of IWAS for prediction of wound assessment class. In current section we describe experiment design comprehensively.

### 1) Data Collection

We collected data of 5 different patient suffering from skin wounds at different time interval by using proposed sensing circuit of IWAS. We collected 150 instances from each patient at different time interval and 5 instances of different interval from each case as shown below in Figure 10.

### 2) Data Preprocessing

Data which we collected from real time wound environment might contain noise, inconsistency, and missing values. Therefore, it is necessary to applied data preprocessing techniques to the collected dataset in order to prepare, it for further decision making. These techniques reconstruct the input dataset into a new modified form which will be more effective for decision making [44].

Most common preprocessing techniques are data discretization and data normalization. We applied data normalization, which used to normalize each attribute values into specified range, like 0.0 to 1.0. There are many methods of normalization like min-max normalization, decimal scaling, and z-score normalization. We applied Min-max to performs a linear transformation on the raw data [45]. It is computed by the formula given in Equation 5.

$$\frac{\ldots MinA}{\ldots - MinA}) * (newMax - newMin) + newMin \quad (5)$$

Where A' contains Min-Max Normalized data

Predefine boundary [newMin, newMax]

| Case No | Time | B.Temp | A.Temp | A.Humidity | Sp2O |
|---|---|---|---|---|---|
| Case 1 | 10:57:24 | 36 | 23 | 20 | 95 |
| | 11:10:27 | 36.2 | 23.5 | 21 | 95.9 |
| | 11:20:30 | 36.5 | 24 | 22 | 95.3 |
| | 12:15:31 | 36 | 23 | 22 | 95 |
| | 13:37:33 | 36.3 | 23.5 | 21 | 95 |
| Case 2 | 11:01:24 | 36.5 | 24 | 80 | 96.2 |
| | 11:10:27 | 36.7 | 24.5 | 81 | 96.3 |
| | 0:20:30 | 36 | 24 | 82 | 96 |
| | 12:50:31 | 36 | 24 | 89 | 96 |
| | 13:37:33 | 37 | 23.5 | 83 | 96 |
| Case 3 | 11:10:48 | 37 | 23 | 90 | 93 |
| | 11:50:51 | 36.2 | 23.5 | 91 | 93 |
| | 0:30:53 | 36.8 | 24 | 91 | 93 |
| | 13:30:56 | 36 | 23 | 92 | 93.3 |
| | 14:10:59 | 36 | 23.5 | 92 | 93 |
| Case 4 | 11:14:59 | 39 | 23 | 20 | 90 |
| | 13:14:02 | 39 | 24 | 19 | 90.2 |
| | 14:14:05 | 38.9 | 24 | 18 | 90.1 |
| | 15:14:08 | 38.9 | 24.5 | 18 | 90.8 |
| | 16:14:12 | 39 | 24.5 | 19 | 90 |
| Case 5 | 10:30:59 | 36 | 24.5 | 22 | 93 |
| | 11:30:02 | 37.5 | 24 | 23 | 92 |
| | 12:35:05 | 37 | 23.5 | 22 | 94 |
| | 13:20:08 | 37 | 23.5 | 22 | 92 |
| | 14:30:12 | 37.5 | 23.5 | 23 | 90 |

**FIGURE 10.** Input Data Set of 5 Cases At Different Time Intervals

**Table 9.**
Decision Tree Predictions

| Case Id | Sample Size | Total Predicted TP | Not Predicted NP | Correctly Predicted CP | Wrongly Predicted WP | Precision TP/TP+WP | Recall TP/TP+NP |
|---|---|---|---|---|---|---|---|
| 1 | 50 | 49 | 1 | 44 | 6 | 89% | 98% |
| 2 | 50 | 50 | 0 | 45 | 5 | 90% | 100% |
| 3 | 50 | 49 | 1 | 47 | 3 | 94% | 98% |
| 4 | 50 | 50 | 0 | 43 | 7 | 87% | 100% |
| 5 | 50 | 50 | 0 | 45 | 5 | 90% | 100% |

### 3) Data Sampling

After collection of input instances, it was necessary to preprocess data in order to convert it meaningful form. As we collected 150 samples from each patient, so we applied random data sampling technique to choose input data cases from datasheets of 5 different patient data consisting 150 instances. We pick 50 input values from each case on random basis in which each input instance having equal chance of selection. [46,47].

### 4) Data Classification Results

We provided sample test data values of each patient case, which we obtained after random selection from 150 instances. We provided these instances to trained decision tree module of proposed IWAS in order to predict assessment class. Table 9 shown obtained prediction results of proposed trained model of decision trees. We used two statistical measures to check performance of decision tree training i.e. Precision and recall.

Precision: We used precision to evaluate correctness of our proposed decision tree module i.e. we check percentage of true predictions out of total obtained predictions for proposed system decision tree. We used formula given in Equation (6) to calculate precision of proposed system.

$$Precision = \frac{Tp}{Tp+Wp} \quad (6)$$

Where Tp=Truly Predicted Instances

Wp=Wrongly Predicted Instances

Recall: We used recall to check completeness of proposed decision tree classification i.e. we measured how much instances are covered by decision tree for classification. We used formula given in Equation (7) to calculate recall of proposed system.

$$Recall = \frac{Tp}{Tp+Np} \quad (7)$$

Where Tp=Truly Predicted Instances

Np=Not Predicted Instances

Figure 11 showed performance graph of obtained results for classification by proposed decision tree, graph showed 87 to 94 % precision for each patient case id 1-5 and recall 98 to 100 % for patient case id 1-5. This obtained percentage for used statistical measure showing that proposed decision tree can efficiently have performed on our provided input cases.

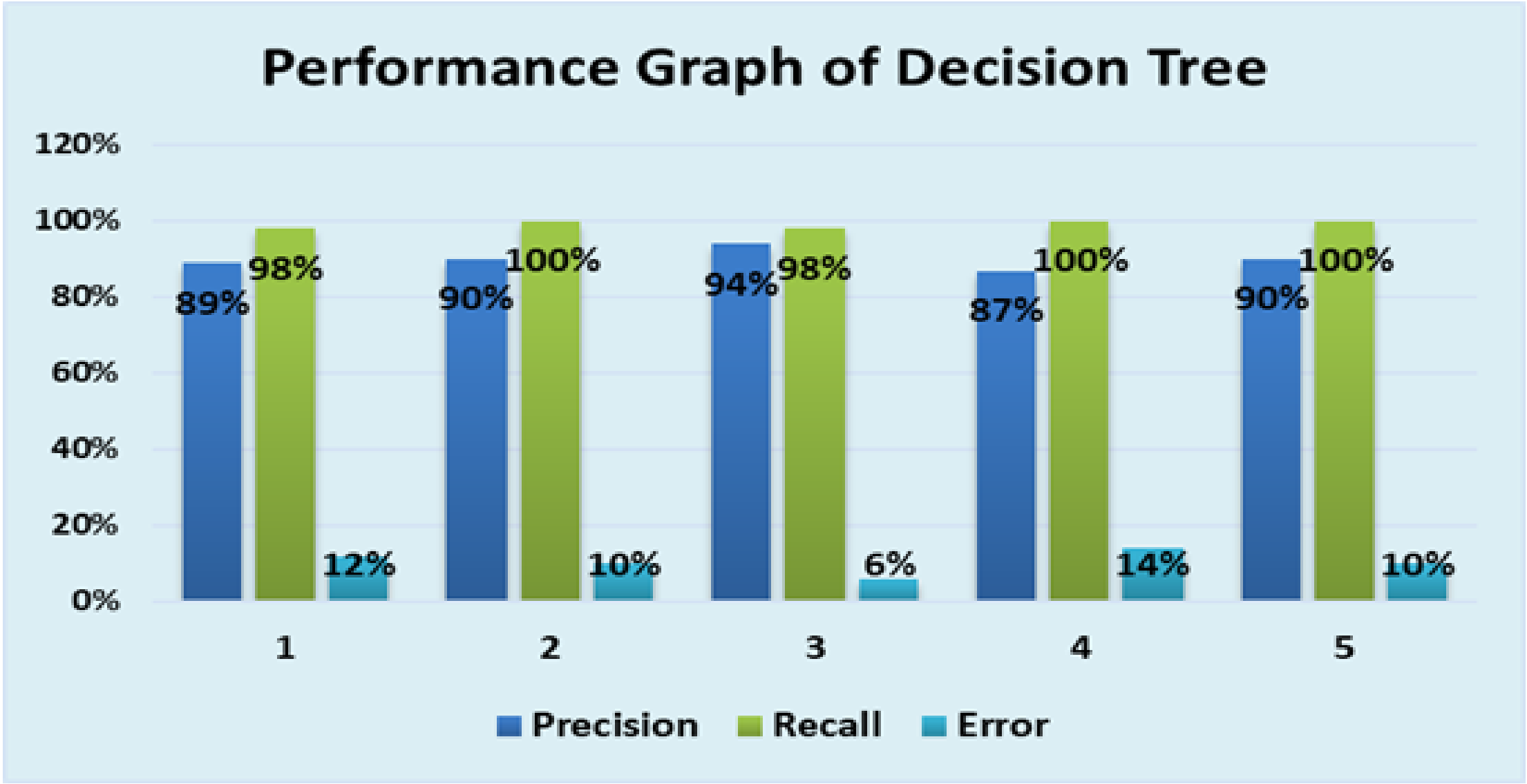

**FIGURE 11.** Data analysis results of Trained Decision Tree

### C. *Limitations of IWAS*

The proposed wound assessment system was proposed to assess wound status in order to analyze healing situation is good, satisfactory or alarming and it was very effective for this purpose. The limitations of proposed solution are described in Table 10.

TABLE 10.
LIMITATIONS OF IWAS

| Concern | Limitation |
|---|---|
| **Technique** | Proposed IWAS used decision tree ,other techniques may also applied for decision making, which could provide better accuracy rate . |
| **Consistency** | Proposed IWAS used training dataset to trained decision tree ,performance of proposed decision tree may vary by changing training data set size and values. |
| **Patient Health Profile** | Proposed IWAS ,assessment of wound is independent of patient health profile i.e. if patient have some other skin diseases or diabetes then proposed system didn't consider them in decision making. |

## VI. Conclusions and Future Work

In clinical research wound assessment techniques usually focused on physical appearance of wound e.g. wound size, wound color, wound shape etc. However other factors also play significant role in healing process therefore they may need to measure for wound assessment. These factors included atmosphere of wound, temperature of wound and amount of oxygen supplied to wound site. These factors can not only influence wound healing but they also showed correlation with other wound characteristic as well i.e. air humidity level correlated with air temperature which in turn effect wound temperature, similarly oxygen levels correlated with wound hydration level and vice versa. For efficient wound assessment system, it is necessary to focus on these factors. In current research we proposed an intelligent wound assessment system based on sensors. Our proposed solution consists of two modules, first module composed of Arduino based circuit, which we used to sense wound factors by using sensors LM35, MAX30100 and DHT22 for wound temperature, oxygenation levels, air temperature and air humidity respectively. In second module we used entropy based decision tree for efficient decision making of wound assessment status based on measurement of wound factors. Our proposed decision tree module classified input in to three possible assessment classes based on their values i.e. good, satisfactory and alarming. Decision tree used entropy and information gain statistics to choose best feature for split. We implemented proposed assessment system in MATLAB. In MATLAB complex tree used entropy and information gain to choose feature for best split, which in turn increased decision tree accuracy rate as results showed that decision tree trained with 96.4 % accuracy rate. We applied proposed IWAS on 5 patient data and got 89-94% precision rate and 98-100% recall, which showed that proposed system produced maximum 94% correctness and 100% completeness rate.
Currently IWAS used decision tree for decision making, but there are many other intelligent data mining techniques which can be used for decision making. So presented IWAS can replicate in future by using different decision making i.e. KNN, fuzzy system, random forest etc. Moreover, there are many other factors which could include in presented system for wound assessment in future i.e. wound infection, wound color etc.